\documentclass[11pt]{evn}
\usepackage[T2A]{fontenc}
\usepackage[cp1251]{inputenc}
\usepackage{epsf}
\def\deg{$^{\circ}$}
\def\ergs{\ifmmode \mathrm{erg\hspace{1mm}s}^{-1} \else erg s$^{-1}$\fi}

\long\def\maintitle#1{{\vskip 20mm \begin{center}\section*{#1}\end{center}\nopagebreak[4]}}

\long\def\author#1{{\begin{center}\normalsize{\bf#1}\end{center}\vskip-1em\index{#1}}\nopagebreak[4]}
\long\def\address#1{{\begin{center}\small\noindent#1\end{center}\vskip-8mm}\nopagebreak[4]}

\begin{document}
\noindent\mbox{\small The 13$^{th}$ EVN Symposium \& Users Meeting Proceedings, 2016}

\maintitle{VLBI STUDIES OF TANAMI RADIO GALAXIES}

\author{R.~Angioni$^{1}$, F. R\"osch$^2$, E.~Ros$^{1,3,4}$,
  M.~Kadler$^{2}$, R.~Ojha$^{5}$, C.~M\"uller$^{6,1}$ and R.~Schulz$^{7}$
  for the TANAMI collaboration}

\address{$^{1}$Max-Planck-Institut f\"ur Radioastronomie, Bonn,
  Germany\\$^{2}$Institut f\"ur Theoretische Physik und Astrophysik,
  Universit\"at W\"urzburg, Germany\\$^{3}$Departament d'Astronomia i
  Astrof\'isica, Universitat de Val\`encia, Spain\\$^{4}$Observatori
  Astron\`omic, Universitat de Val\`encia, Spain\\$^{5}$NASA Goddard
Space Flight Center, Greenbelt MD, USA\\$^{6}$Department of
Astrophysics/IMAPP, Radboud University Nijmegen, the Netherlands\\$^{7}$ASTRON, The Netherlands Institute for Radio
Astronomy, Dwingeloo, the Netherlands}

\begin{abstract}

  Radio galaxies are relatively faint at $\gamma$-ray energies, where they make up only 1--2\%
  of all AGN detected by {\it Fermi}-LAT. However, they
  offer a unique perspective to study the intrinsic properties of AGN
  jets. For this reason, the combination of $\gamma$-ray and multi-wavelength data with
  high-resolution VLBI monitoring is a powerful tool to tackle the
  basic unanswered questions about AGN jets. Here we present
  preliminary results from a sample study of radio galaxies in the
  Southern hemisphere observed by the TANAMI VLBI monitoring program. We obtain
  high-resolution maps at 8.4 and 22.3 GHz, and study the jet kinematics
  using multi-epoch data. We present a preliminary kinematic
  analysis for the peculiar $\gamma$-ray AGN PKS~0521$-$36.

\end{abstract}

\section{Introduction}
AGN dominate the extragalactic high-energy $\gamma$-ray sky, as revealed
by the Third {\it Fermi}-LAT Gamma-ray source List (3FGL, \cite{3fgl}).
The great majority of the sources are blazars, i.e. radio-loud AGN
with relativistic jets oriented at small angles to the observer's
line-of-sight. Their radiation is strongly
Doppler boosted and shows fast variability. Blazars are among the brightest sources of radiation in
the universe, spanning the whole electromagnetic spectrum, from radio
to $\gamma$-rays. Relativistic jets have been studied for half a century, but their
fundamental inner workings are still not well understood, e.g. their
acceleration and collimation mechanisms, or the location of the
$\gamma$-ray emission region.

Radio galaxies are believed to be the parent population of blazars, with jets
oriented at larger angles \cite{urry95}. Because of their misaligned orientation,
their radiation is much less Doppler boosted with respect to blazars.
Therefore, radio galaxies are typically orders of magnitude fainter than their
aligned counterpart, across the whole spectrum. Because of the limited
sensitivity of $\gamma$-ray instruments, this has particularly evident
implications at high-energies. Indeed, {\it Fermi}-LAT has detected
less than 20 radio galaxies so far, while the 3FGL includes more than
1000 associated blazars \cite{3lac}. In spite of the small sample size and lower fluxes, which
complicate radio galaxy studies at high-energies, they provide us with
a view on $\gamma$-rays from AGN jets which is less biased by
relativistic effects, and is complementary to blazar studies \cite{magn}.

The combination of $\gamma$-ray data with high-resolution radio
observations can be a powerful tool in investigating these questions.
{\it Fermi}-LAT has an energy-dependent angular resolution of the order of $\sim$0.1\deg, therefore almost all
AGN observed in $\gamma$-rays appear as point sources, and it is not
possible to directly associate the high-energy emission to a specific
morphological component. VLBI observations, on the other hand, are able to
achieve milliarcsecond resolutions. Since AGN are variable sources,
it is possible to identify the $\gamma$-ray emission region on VLBI
scales by looking for correlated variability in radio morphology or
flux, and high-energy emission (e.g. \cite{cas15,kar16}). Additionally, multi-epoch VLBI observations
provide the only direct measure of relativistic jet motion, allowing
us to derive relevant jet parameters such as apparent speed and jet
orientation angle.

\section{The TANAMI monitoring program of Southern-hemisphere AGN}
TANAMI (Tracking Active galactic Nuclei with Austral Milliarcsecond
Interferometry) is a multi-wavelength monitoring program of AGN jets south of
$-30^{\circ}$ declination, including almost $\sim100$ sources. The sample
was defined as a hybrid radio flux-limited sample combined with
sources detected in $\gamma$-rays or reported as likely high-energy
emitters \cite{tanami}. 

The core of the program is VLBI monitoring at 8.4 GHz and 22.3 GHz,
since 2007. The array is based on the Australian Long Baseline Array (LBA),
supported by associated antennas in South Africa, New Zealand,
Antarctica and Chile (a full list of the participating antennas can be
found in \cite{mul14}). First-epoch 8.4 GHz images at milliarcsecond resolution were presented for an initial sample of 43 sources
by \cite{tanami}. The sample was constantly expanded with new
$\gamma$-ray detections or other interesting new sources, and
first-epoch images of the first-ever high-resolution observations, for
most sources, will be presented in a
forthcoming paper (M\"uller et al. in prep.). The radio VLBI monitoring is
complemented by an excellent multi-wavelength coverage, including NIR, optical, UV, X-ray and
$\gamma$-ray data, providing a quasi-simultaneous broadband view of
the sources, as is required for detailed studies of variable sources
such as AGN. An overview of the multi-wavelength program and selected
TANAMI results can be found in \cite{kad15}.

\section{Radio galaxies in TANAMI}
Since the TANAMI sample was selected based on current or expected
$\gamma$-ray detections, the majority of the sources are blazars, but
several well-known radio galaxies are also monitored. An example is
the closest radio-loud AGN, Centaurus A, for
which TANAMI data revealed the complex dynamics of the jet's inner
parsec, indicating downstream jet acceleration \cite{mul14} and a spectral
morphology \cite{mul11} that can be explained within the spine-sheath scenario \cite{tav08}. 

The TANAMI sample includes fifteen additional
radio galaxies, including several notable sources for which TANAMI is
able to provide the highest resolution data available: the classic FR II Pictor A, whose jet has
been detected and resolved in X-rays \cite{wil01} and optical \cite{gen15} as well
as radio bands \cite{tin00}; the FR I PKS~0625$-$35, which is the most
recent addition to the elusive group of only 5 radio galaxies to be detected in the TeV band by
Cherenkov telescopes \cite{dyr15}, and shows evidence of superluminal
motion in a preliminary kinematic analysis of TANAMI data \cite{mul12}; the nearby FR I Centaurus~B,
detected in $\gamma$-rays by {\it Fermi}-LAT with a notably flat spectral index \cite{3lac};
the peculiar object PKS~0521$-$36,
showing $\gamma$-ray variability and originally classified as a BL Lac
but later shown to have properties consistent with a larger viewing
angle \cite{dam15}. The full list of TANAMI radio galaxies is given in
Table~1.

The multi-wavelength properties of these sources are currently under
study within the TANAMI collaboration (Angioni et al. in prep.), including images, kinematic analysis and spectral
index maps. 

\section{The peculiar $\gamma$-ray AGN PKS~0521$-$36}
Here we present preliminary
results on one $\gamma$-ray bright TANAMI radio galaxy, PKS~0521$-$36. This is a nearby ($z=0.0565$)
AGN with uncertain classification. Leon et al. 2016 \cite{leo16} classify it as a BL
Lac, and derive limits on the jet viewing angle, speed, and Doppler
factor using the Atacama Large Millimeter Array (ALMA). Their results suggest a jet viewing angle in
the range
16\deg$\le\theta\le$38\deg. D'Ammando et al. 2015 \cite{dam15} constrain the same parameters
using SED modeling including $\gamma$-ray data, obtaining a more
aligned jet viewing angle of 6\deg$\le\theta\le$15\deg. These results
point to an intermediate jet viewing angle between a blazar and a
steep spectrum radio quasar (SSRQ) or radio galaxy.

Previous VLBI observations performed with the VLBA and with the Southern Hemisphere
VLBI Experiment (SHEVE) at 4.9 GHz and 8.4 GHz provided an upper limit on the apparent speed of
jet components $\beta_{app}<1.2$ \cite{tin02}. This is also consistent
with the hypothesis that the jet of PKS~0521$-$36 is not strongly beamed.
The authors mentioned that future more sensitive VLBI data would be
able to provide more stringent constraints on the nature of this
source.

\begin{figure}[!htbp]
{\epsfclipon\epsfxsize=0.45\textwidth  \epsffile{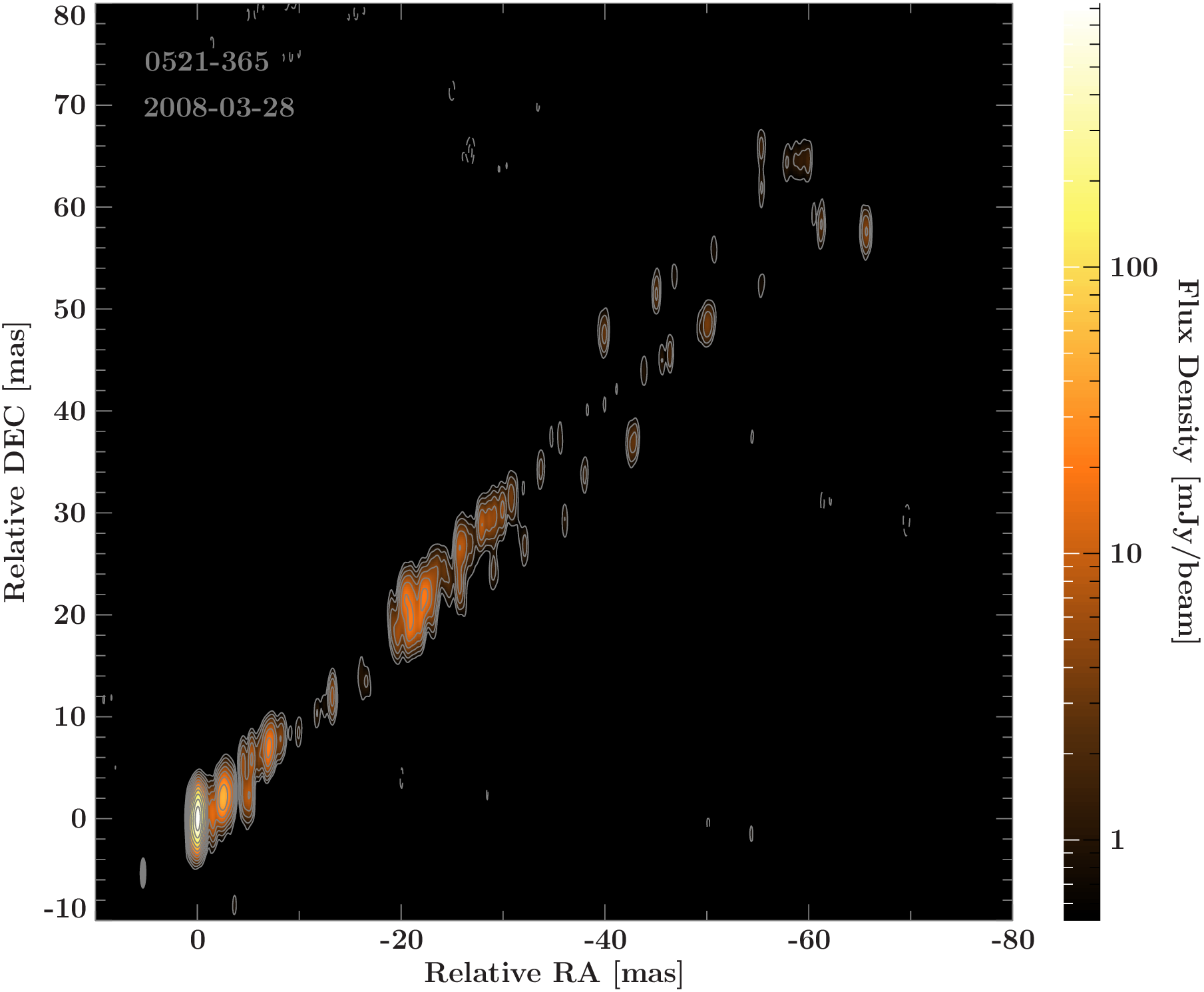}}
{\epsfclipon\epsfxsize=0.55\textwidth \epsffile{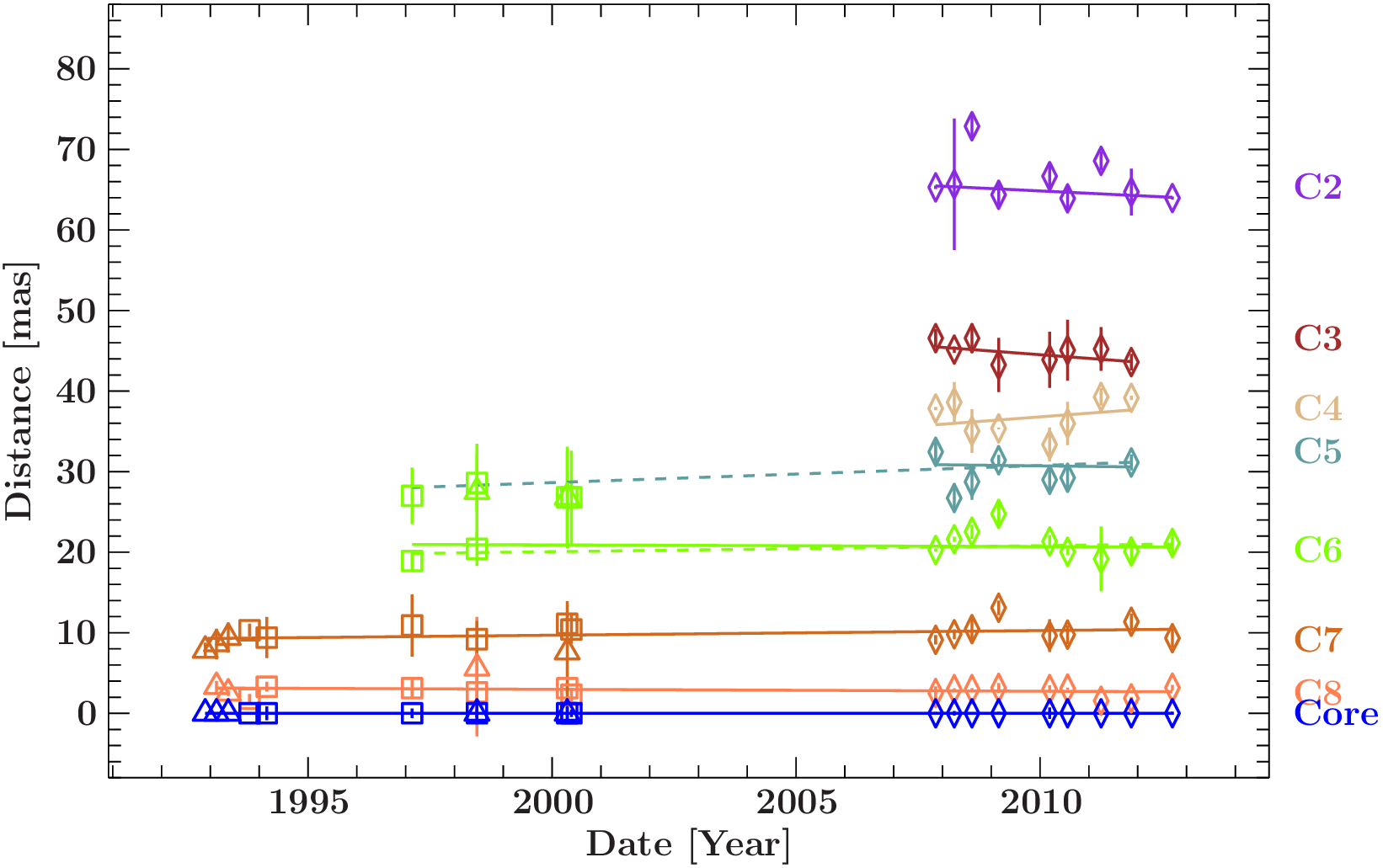}}
\
\caption{{\it Left panel}: TANAMI full-resolution 8.4 GHz images of PKS~0521$-$36. The contours start
  at 3 times the rms level. The beam size is indicated in grey in the
  lower left corner. {\it Right panel}: Results of the preliminary kinematic analysis for PKS~0521$-$36, including TANAMI data and previous VLBI data from
  \cite{tin02}. The plot shows the distance of jet component relative
  to the core as a function of time. Triangles and squares indicate
  4.9 GHz data and 8.4 GHz data from \cite{tin02}, respectively.
  Diamonds indicate TANAMI 8.4 GHz data. The errors are estimated as half the
  component major axis.}
\label{maps}

\end{figure}

The TANAMI 8.4 GHz full-resolution image presented in the left panel of Fig.~\ref{maps} achieves an order of
magnitude improvement in sensitivity with respect to previous VLBI
data \cite{tin02}, allowing to reveal a more extended jet\footnote{A
  first-epoch map of PKS~0521$-$36 was already presented in \cite{tanami}.}. TANAMI
monitoring provided 9 epochs for this source between 2007 and
2013, an excellent data set for kinematic analysis. Preliminary
results from this analysis are shown in the right panel of Fig.~\ref{maps}, where the
distance of jet components from the core is plotted as a function of
time. The data shown include all TANAMI epochs and the previous VLBI
data from \cite{tin02}, providing kinematic information across a
${\sim}20$ years time span. A linear regression fit is applied to the
components, with two of them being cross-identified between the two data
sets, yielding an estimate of their apparent speed. 

This preliminary analysis confirms the absence of fast
jet apparent motions in PKS~0521$-$36, with the largest possible
apparent speed (for the components that can be reliably identified and
fitted) being $\beta_{app}\sim0.36$ for component C7 (see right panel
of Fig.~\ref{maps}). This, together with
estimates of the jet-to-counterjet ratio, constrains the intrinsic jet
speed and viewing angle to a narrower region of the parameter space
with respect to previous studies, namely $\beta>0.56$ and
$\theta<16$\deg. This viewing angle estimate agrees with the estimate
provided by \cite{dam15}, supporting the hypothesis that the jet of
PKS~0521$-$36 is oriented at an intermediate angle to the line-of-sight.


\begin{table}[htbp]
\label{rgs}

\caption{Radio galaxies in the TANAMI sample.}
\vspace{0.25cm}
\centering

\begin{footnotesize}

\begin{tabular}{@{}llllc@{}}
\hline
\hline
Source & Catalog & Class & $z$ & {\it Fermi}-LAT det.\\
\hline
0518$-$458 & Pictor~A & FR II & 0.035 & yes\\
0521$-$365 & PKS~0521$-$36 & RG/SSRQ & 0.0565 & yes\\
0625$-$354 & PKS~0625$-$35 & FR I/BLL & 0.0546 & yes\\
1258$-$321 & PKS~1258$-$321 &  FR I & 0.017 & no\\
1322$-$428 & Centaurus~A & FR I & 0.0018 & yes\\
1333$-$337 & IC 4296 &  FR I & 0.0125 & no\\
1343$-$601 & Centaurus~B &  FR I & 0.0129 & yes\\
1549$-$790 & PKS~1549$-$79 &  RG/CFS & 0.150 & no\\
1600$-$489 & PMN~J1603$-$4904 & MSO$^a$ & 0.18 & yes\\
1718$-$649 & NGC~6328 &  GPS/CSO & 0.0144 & yes\\
1733$-$565 & PKS~1733$-$56 &  FR II & 0.098 & no\\
1814$-$637 & PKS~1814$-$63 &  CSS/CSO & 0.0627 & no\\
1934$-$638 & PKS~1934$-$63 & GPS & 0.18 & no\\
2004$-$447 & PKS~2004$-$447 & NLSy1/CSS$^b$ & 0.24 & yes\\
2027$-$308 & PKS~2027$-$308 &  RG & 0.539 & no\\
2152$-$699 & PKS~2153$-$69 &  FR II & 0.0283 & no\\
\hline
\hline
\end{tabular}
\begin{flushleft}
$^a$ Classified as a young radio galaxy based on multi-wavelength
studies \cite{mul14b,mul15,mul16}.\\
$^b$ Evidence for classification as young radio galaxy \cite{kre16,sch16}.
\end{flushleft}
\end{footnotesize}

\end{table}




\end{document}